# Fabrication and characterization of InSb nanosheet/hBN/graphite heterostructure devices

Li Zhang[1], Yuanjie Chen[1], Dong Pan[2], Shaoyun Huang[1], Jianhua Zhao[2], and H. Q. Xu[1,3,*]

[1]*Beijing Key Laboratory of Quantum Devices, Key Laboratory for the Physics and Chemistry of Nanodevices, and School of Electronics, Peking University, Beijing 100871, China*

[2]*State Key Laboratory of Superlattices and Microstructures, Institute of Semiconductors, Chinese Academy of Sciences, P.O. Box 912, Beijing 100083, China*

[3]*Beijing Academy of Quantum Information Sciences, Beijing 100193, China*

[*]Corresponding author: hqxu@pku.edu.cn

(Date: April 18, 2022)

## Abstract

Semiconductor InSb nanosheet/hexagonal boron nitride (hBN)/graphite trilayers are fabricated, and single- and double-gate devices made from the trilayers are realized and characterized. The InSb nanosheets employed in the trilayer devices are epitaxially grown, free-standing, zincblende crystals and are in micrometer lateral sizes. The hBN and graphite flakes are obtained by exfoliation. Each trilayer is made by successively stacking an InSb nanosheet on an hBN flake and on a graphite flake using a home-made alignment stacking/transfer setup. The fabricated single- and double-gate devices are characterized by electrical and/or magnetotransport measurements. In all these devices, the graphite and hBN flakes are employed as the bottom gates and the gate dielectrics. The measurements of a fabricated single bottom-gate field-effect device show that the InSb nanosheet in the device has an electron field-effect mobility of ~ 7300 $cm^2$ $V^{-1}$ $s^{-1}$ and a low gate hysteresis of ~ 0.05 V at 1.9 K. The measurements of a double-gate Hall-bar device show that both the top and the bottom gate exhibit strong capacitive couplings to the InSb nanosheet channel and can thus tune the nanosheet channel conduction effectively. The electron Hall mobility in the InSb nanosheet of the Hall-bar device is extracted to be larger than $1.1 \times 10^4$ $cm^2$ $V^{-1}$ $s^{-1}$ at a sheet electron density of ~ $6.1 \times 10^{11}$ $cm^{-2}$ and 1.9 K and, thus, the device exhibits well-defined Shubnikov-de Haas oscillations.

# 1. Introduction

Because of their small electron effective mass, strong spin-orbit interaction (SOI), and large Landé *g*-factor, low-dimensional InSb nanostructures are attractive candidates for applications in high-speed electronics [1,2], spintronics [3,4], quantum electronics [5-7], and topological quantum computation [8,9]. To date, epitaxially grown InSb nanowires have been extensively investigated and have been employed as emerging platforms for realizing field-effect transistors (FETs) [1,2], quantum dot devices [5-7], and semiconductor-superconductor hybrid quantum devices [10-12]. Recently, high-quality two-dimensional (2D) InSb quantum structures, such as InSb quantum wells (QWs) [13-21] and free-standing nanosheets [22-25], have been achieved via epitaxial growth, opening up new possibilities for fabricating planar, integrated, complex quantum devices. Due to advances in material engineering, these heterostructured InSb QWs have been demonstrated to exhibit high electron mobilities and large mean free paths, and have been successfully employed to make good quantum point contacts (QPCs) in which well-defined conductance quantization has been observed [14,17,21]. In comparison, an advantage of using free-standing InSb nanosheets to prepare devices is that Ohmic or superconducting contacts can be made by directly depositing suitable metals on their surfaces. Currently, InSb nanosheets have been used for fabricating Hall-bar devices [22,24,25], QPCs [23], Josephson junctions (JJs) [26-28], quantum dot devices [29,30], and dual-gate field-effect devices [31]. However, most of these devices are fabricated on $SiO_2$/Si substrates, and scattering by charge traps and defects in the $SiO_2$ dielectrics and at the interfaces between the InSb nanosheets and the $SiO_2$ dielectrics have limited the device performance and led to a difficulty in achieving high-quality quantum devices.

One possible solution to the above problem is to adopt a new layered, charge-trap free material with an atomically flat surface as a dielectric and a thin graphite as a gate electrode [32-35]. In the past few years, hexagonal boron nitride (hBN) has been proven to be such an excellent dielectric [36]. With their ultra-clean surfaces, fresh exfoliated hBN layers have been assembled, together with a variety of one-dimensional (1D) and 2D materials, such as graphene [36,37], transition metal dichalcogenides (TMDCs) [33,38-40], black phosphorus [32,41,42], and InSb nanowires [43,44], into functional devices. In these devices, carrier mobilities have been significantly increased, when compared to their counterpart devices fabricated directly on $SiO_2$/Si substrates, and quantum oscillations and ballistic transport at low magnetic fields [43] have been observed. Therefore, it is desirable to transfer InSb nanosheets onto hBN/graphite bilayers to construct devices with hBN as a dielectric and graphite as a gate electrode on $SiO_2$/Si substrates in order to suppress the influence of charge traps and defects at the surfaces of and inside the $SiO_2$ dielectrics.

In this article, a method of fabrication of free-standing micrometer-sized InSb nanosheet/hBN/graphite trilayers is described, and realization of single- and double-gate devices from the trilayers on $SiO_2$/Si substrates is demonstrated. As a key step of the method, the positions of almost optically invisible InSb nanosheets are determined with use of duplicated patterns on the transparent stacking/transfer stamps and thus tiny InSb nanosheets can be optically aligned to and stacked on exfoliated hBN flakes and then on exfoliated graphite flakes. In the fabricated single- and double-gate devices, the graphite and hBN flakes are used as the bottom gates and the gate dielectrics. The devices are characterized by electrical and/or magnetotransport measurements. In this work, typical results obtained only from the gate-transfer characteristic measurements of a single-gate field-effect device [as shown in Figs. 2(b) and 2(c)] and from the electrical and magnetotransport measurements of a double-gate Hall-bar device [as shown in Figs. 3(b) and 3(c)] will be presented. The gate transfer characteristic measurements of the single-gate field-effect device yield an electron field-effect mobility of ~7300 $cm^2\,V^{-1}\,s^{-1}$ in the InSb nanosheet and a low hysteresis of ~0.05 V in gate voltage at low temperatures. The electrical and magnetotransport measurements of the double-gate Hall-bar device show that the two gates exhibit strong capacitive couplings to the InSb nanosheet channel and the electron Hall mobility of the InSb nanosheet becomes more than $1.1 \times 10^4\,cm^2\,V^{-1}\,s^{-1}$ at a sheet electron density of ~ $6.1 \times 10^{11}\,cm^{-2}$ and 1.9 K, leading to observation of well-defined Shubnikov-de Haas oscillations.

## 2. Experimental details

The InSb nanosheet/hBN/graphite trilayer heterostructures used in the single-gate field-effect InSb nanosheet device and the double-gate InSb nanosheet Hall-bar devices studied in this work are obtained by mechanical alignment/transfer technique. The details about the trilayer heterostructure fabrication is given in Supplementary Information. Here, taken as an example, only a brief description about the fabrication of an InSb nanosheet/hBN/graphite trilayer heterostructure shown in Fig. 1(d) is presented. The InSb nanosheet used in the fabrication of the trilayer heterostructure was grown in a free-standing form on InAs nanowire stems via molecular beam epitaxy (MBE) on a p-type Si(111) substrate [22] and was transferred from the growth substrate to a $SiO_2$/Si substrate with predefined Ti/Au metal markers (referred to as sample substrate A). Figure 1(a) displays an SEM image of the InSb nanosheet and Ti/Au markers on sample substrate A. Here, the InSb nanosheet is located at the upper side of marker '8', as indicated by a black dashed square. The nanosheet had a length of ~ 1 μm and a width of ~ 0.5 μm. The inset of Fig. 1(a) shows a high-resolution transmission electron microscope (HRTEM) image of a typical MBE-grown InSb nanosheet taken from the same growth substrate and the corresponding fast Fourier transform (FFT) pattern (graph in the lower-right corner of the inset). Here, it is shown that the nanosheet was a zinc-blende crystal, and free from stacking faults and twin defects. The hBN and graphite flakes were mechanically exfoliated from commercially available bulk hBN and graphite materials using Scotch tapes and were transferred onto two other marker-prepared $SiO_2$/Si substrates (referred to as sample substrates B and C).

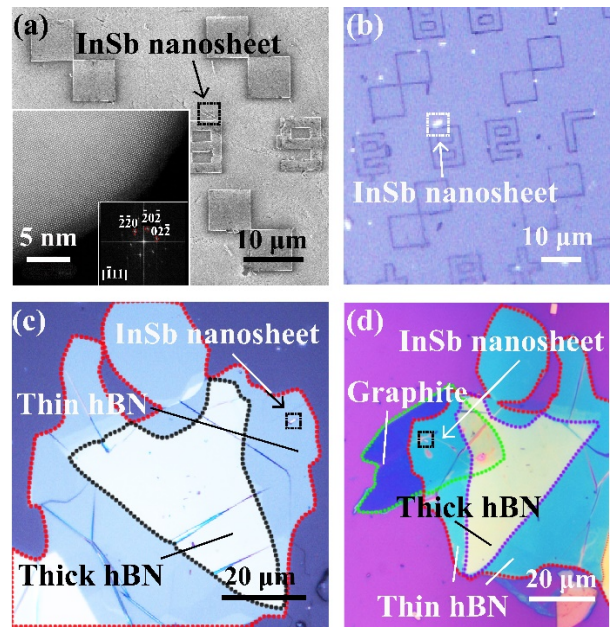

**Figure 1.** SEM and optical images of a micrometer-sized InSb nanosheet/hBN/graphite trilayer sample at different fabrication stages. (a) SEM image of an InSb nanosheet and Ti/Au markers on a $SiO_2$/Si substrate. The nanosheet is located in the region indicated by a black dashed square. Inset: HRTEM image of a typical MBE-grown InSb nanosheet and the corresponding FFT pattern (graph in the lower-right corner of the inset). (b) Optical image of the PPC surface of the stamp, on which the same InSb nanosheet as in (a) and the duplicated patterns of the substrate markers are seen. The InSb nanosheet is in the region indicated by a white dashed square. (c) Optical image of the InSb nanosheet/hBN bilayer after a hBN flake being picked up. The InSb nanosheet is located in the area indicated by a black dashed square. The light blue region marked by a red dotted line and the white region marked by a black dotted line are the pieces of thin and thick hBN. (d) Optical image of the finally fabricated InSb nanosheet/hBN/graphite trilayer. The InSb nanosheet is located in the area indicated by a black dashed square. The thick hBN, thin hBN, and graphite regions are marked by purple, red, and green dotted lines, respectively.

To fabricate the InSb nanosheet/hBN/graphite trilayer heterostructure, the InSb nanosheet was first picked up from substrate A and was then stacked on the hBN flake on substrate B to form an InSb nanosheet/hBN bilayer using a home-made alignment stacking/transfer setup mounted with a polypropylene carbonate (PPC) film coated polydimethylsiloxane (PDMS) stamp [34]. The InSb nanosheet/hBN bilayer was then picked up by lifting the stamp from substrate B. Here, it is important to note that to align the almost optically invisible, micrometer-sized InSb nanosheet with the hBN flake, a pattern of the optically visible metal markers on substrate A was duplicated on the PPC film during the nanosheet capture and was used for the material alignment. It should also be noted that before the stamp was lifted, the stamp was pressed hard against substrate B, in order for the InSb nanosheet and the hBN flake to form a tightly attached bilayer on the PPC film of the stamp. Figure 1(b) shows an optical image of the same InSb nanosheet as in Fig. 1(a) after being captured on the PPC film, on which the duplicated pattern is also clearly seen. The InSb nanosheet lay at the upper side of duplicated pattern '8', as indicated by a white dashed square. Figure 1(c) shows an optical image of the InSb nanosheet/hBN bilayer on the PPC surface after the bilayer was picked up. Here, it is seen that the entire hBN flake is not in the same thickness and the regions corresponding to the pieces of thick hBN and thin hBN are marked by black and red dotted lines, respectively. The InSb nanosheet as seen in Figs. 1(a) and 1(b) is now found below the thinner portion of the hBN flake as indicated by a black dashed square. Next, the InSb nanosheet/hBN bilayer on the stamp was aligned and stacked on the graphite flake on substrate C. After pressing hard the stamp against substrate C, the stamp was released by heating up sample substrate C, leaving the PPC film coated InSb nanosheet/hBN/graphite trilayer on sample substrate C. Finally, the PPC film was removed from the trilayer by dipping sample C into acetone. Figure 1(d) shows an optical image of the finally achieved trilayer heterostructure on sample substrate C, where the InSb nanosheet is located on top of the hBN/graphite bilayer in a region marked by a black dashed square. For further details about the procedure of making such an InSb nanosheet/hBN/graphite trilayer heterostructure, please refer to Supplementary Information.

The single-gate field-effect device as shown in Fig. 2(b) and the double-gate Hall-bar device as shown in Fig. 3(b) are made, respectively, from an InSb nanosheet/hBN/graphite trilayer heterostructure as shown in Fig. 2(a) and an InSb nanosheet/hBN/graphite trilayer heterostructure as shown in Fig. 3(a), obtained in the process described above. In both devices, the graphite flake and the hBN flake were employed as the bottom gate and the gate dielectric, respectively. In fabrication of the single-gate device, the InSb nanosheet/hBN/graphite heterostructure was first located relative to the predefined markers on the substrate (substrate C) by an optical microscope and some smaller markers were fabricated around the InSb nanosheet by electron-beam lithography (EBL), electron-beam evaporation (EBE) of Ti/Au (5/45 nm in thickness), and lift-off process. Subsequently, the InSb nanosheet was located relative to these small markers by SEM, and source and drain contact electrodes and gate electrodes were fabricated on the InSb nanosheet and on the graphite flake via EBL, EBE of Ti/Au (10/100 nm in thickness), and lift-off. It should be noted that before the metal evaporation, exposed areas on the InSb nanosheet were chemically etched in a de-ionized water-diluted $(NH_4)_2S_x$ solution to remove the surface oxide. In fabrication of the double-gate Hall-bar device, a top gate was made on an otherwise completed single bottom-gate device. This was done first by depositing a dielectric layer of $HfO_2$ (20 nm in thickness) on the sample by atomic layer deposition and then by fabricating a Ti/Au (5/120 nm in thickness) metal gate on top via the combined process of EBL, EBE, and lift-off.

Figures 2(b) and 2(c) show an SEM image of the single-gate field-effect device and a schematic for the layer structure of the device and the electrical measurement circuit setup. The gate-transfer characteristics of the device were measured in a two-probe configuration by applying a fixed DC voltage ($V_{ds}$) of 0.4 mV to source and drain electrodes (i.e., contacts 2 and 3) and detecting the channel current ($I_{ds}$) as a function of bottom-gate voltage $V_{BG}$. Figure 3(b) shows an SEM image of the double-gate Hall-bar device and the magnetotransport measurement circuit setup. Figure 3(c) shows a schematic view of the layer structure of the device and the circuit setup for application of the top and bottom gate voltages. The magnetotransport measurements of the device were carried out using a standard current-biased lock-in technique, in which a 17-Hz AC current ($I$) of 50 nA was supplied between the source and drain electrodes (i.e., contact electrodes 1 and 6), and the voltage drop $V_x$ between probe electrodes 2 and 3 and the voltage drop $V_y$ between probe electrodes 5 and 3 were recorded. The longitudinal and Hall resistances were then obtained numerically from $R_{xx} = V_x/I_x$ and $R_{yx} = V_y/I_x$. All the measurements were carried out in a physical property measurement system cryostat equipped with a uniaxial magnet and, in the magnetotransport measurements for the double-gate Hall-bar device, the magnetic field was applied perpendicular to the InSb nanosheet plane.

## 3. Results and discussion

*3.1. Gate-transfer characteristic measurements of the single-gate field-effect device*

Figure 2 shows the gate-transfer characteristic study of the single-gate field-effect device. Figures 2(a) and 2(b) show an optical image of an InSb nanosheet/hBN/graphite trilayer heterostructure (before fabricating small markers) and an SEM image of the single bottom-gate field-effect device made from the trilayer heterostructure [see Fig. 2(c) for a cross-sectional view of the layer structure of the device]. Note that in preparation of this trilayer heterostructure, hBN flakes were successively picked up twice. This is because, due to misalignment, the target InSb nanosheet was not stacked on the hBN flake picked up at the first time. Thus, in Fig. 2(a), two portions of thick hBN, a big piece of thin hBN (with a thickness of $t$~30 nm), and a piece of graphite are found and are marked by purple, green, and red dotted lines, respectively. The area where the InSb nanosheet is located is indicated by a white square. The InSb nanosheet in the single-gate field-effect device has a width of $W$~800 nm and a thickness of ~20 nm (estimated based on the calibrated contrast in the SEM image). Note that an InSb nanosheet with a relatively thin thickness was chosen here because it is expected that this thickness would be smaller than the Fermi wavelength of the electron and thus the electron transport in the nanosheet is of a 2D nature. The InSb nanosheet was contacted by four Ti/Au electrodes, as shown in Fig. 2(b), where the distance between the two inner contact electrodes is $L$~600 nm.

Figure 2(d) shows the two-terminal conductance $G = I_{ds}/V_{ds}$ of the device as a function of the gate voltage $V_{BG}$ measured at $V_{ds}$=0.4 mV and $T = 1.9$ K (black curve) by sweeping $V_{BG}$ upward [cf. the measurement circuit setup shown in Fig. 2(c)]. The measurements show that the device is an n-type transistor and the InSb conduction channel is in a pinch-off state at $V_{BG} = 0$ V and can be switched on by applying a positive $V_{BG}$. The channel conductance $G_s$ is related to the charge carrier mobility $\mu$ via [1]

$$G_s = \frac{\mu}{L^2} C_g (V_{BG} - V_{th}), \tag{1}$$

where $V_{th}$ is the threshold voltage at which the channel current is cut off, and $C_g$ is the gate-to-channel capacitance which can be estimated from $C_g = \varepsilon_0 \varepsilon_r \frac{LW}{t}$ with $\varepsilon_0$ being the permittivity of vacuum, $\varepsilon_r = 2.9$ the relative dielectric constant of hBN, and $t = 30$ nm the thickness of hBN. The measured resistance $G^{-1}$ is composed of two terms, the channel resistance $G_S^{-1}$ and the contact resistance $R_c$ [1, 45], and is given by,

$$G^{-1} = G_S^{-1} + R_c. \tag{2}$$

The field-effect carrier mobility $\mu$, contact resistance $R_c$, and threshold voltage $V_{th}$ of the device at $T = 1.9$ K can be extracted by fit of the measured on-state transfer characteristic curve shown in Fig. 2(d) to Eqs. (1) and (2). The red curve in Fig. 2(d) shows the result of the fit. The fit yields the values of $\mu$~7300 cm$^2$ V$^{-1}$ s$^{-1}$, $R_c$~580 Ω, and $V_{th}$~0.88 V. Note that the extracted mobility value in the nanosheet is much smaller than the values measured in InSb QWs (51000 cm$^2$ V$^{-1}$ s$^{-1}$ at 300 K, and 248000 cm$^2$ V$^{-1}$ s$^{-1}$ at 4.5 K) [46]. This could be attributed to several factors. The most important one is due to scattering by charge traps and defects at the surfaces and in the native oxidized layers of the nanosheet. Another one is that the extracted value of $\mu$ for the nanosheet is the field-effect mobility which represents an average value over a range of the gate voltages (or carrier densities) and could be much smaller than the Hall mobility values obtained at high carrier densities.

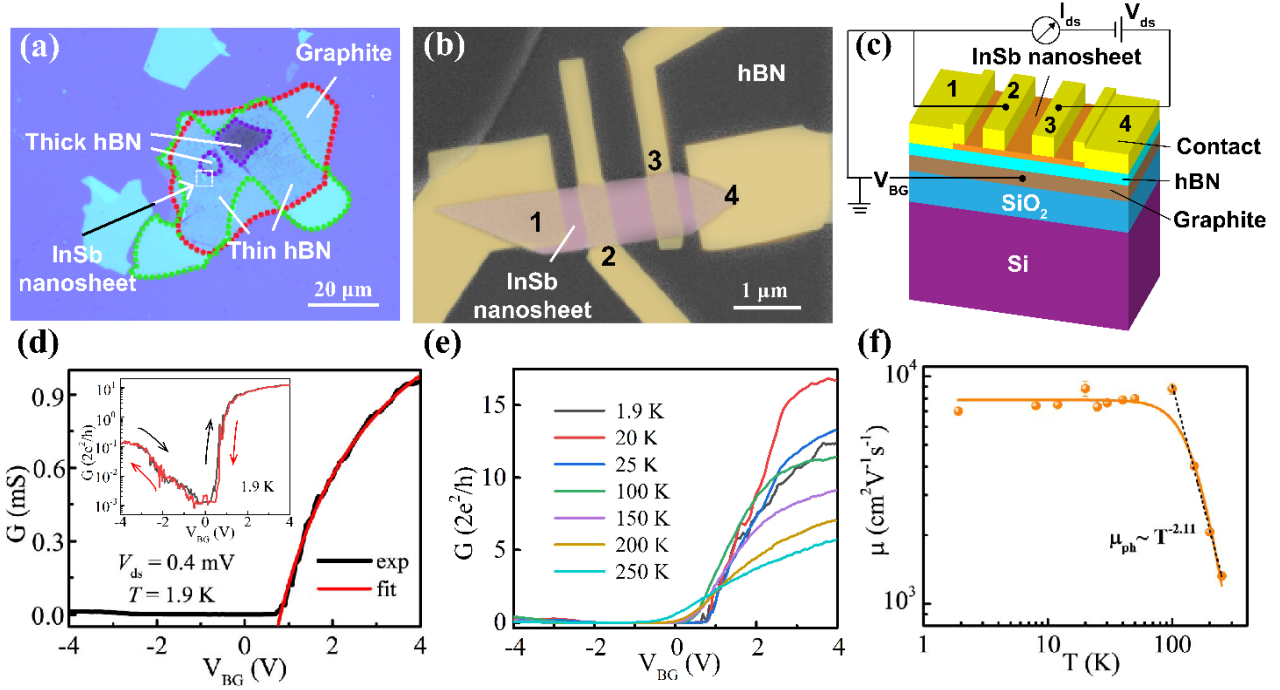

**Figure 2.** Electrical characterization measurements of a single-gate field-effect device. (a) Optical image of the InSb nanosheet/hBN/graphite heterostructure used for fabrication of the device. The InSb nanosheet is located in the area indicated by a white square. Purple, green and red dotted lines mark the thick hBN, thin hBN, and graphite regions in the sample, respectively. (b) False-colored SEM image of the device. The nanosheet has a width of ~800 nm and a thickness of ~20 nm. The separation between the two inner electrodes is ~600 nm. (c) Schematic for the layer structure of the device and the electrical measurement circuit setup. (d) Measured conductance $G = I_{ds}/V_{ds}$ (black curve) as a function of gate voltage $V_{BG}$ (gate-transfer characteristics) for the device at $V_{ds}$=0.4 mV and $T = 1.9$ K by sweeping $V_{BG}$ upward. The red curve shows the fit of the on-state portion of the measured gate-transfer characteristic curve to Eqs. (1) and (2). The inset shows the measured gate-transfer characteristics of the device in a logarithmic scale. The black and red curves represent results of the measurements by sweeping $V_{BG}$ upward and downward, respectively. (e) Gate-transfer characteristics measured at different temperatures $T$. (f) Extracted field-effect mobility $\mu$ as a function of $T$. The orange solid line is the fit of the mobility data at different $T$ to Eq. (3). The black dashed line is the fit of the high-temperature mobility data by a power law $\mu \sim T^{-\gamma}$ with a value of $\gamma \sim 2.11$.

With use of the extracted value of $V_{\text{th}}$, the electron density in the InSb nanosheet at a given value of $V_{\text{BG}}$ can be estimated from $n = C_{\text{gs}} \times \frac{V_{\text{BG}} - V_{\text{th}}}{e}$, where $C_{\text{gs}} = \frac{\varepsilon_0 \varepsilon_r}{t}$ is the unit area capacitance of the graphite gate to the InSb nanosheet and $e$ is the elementary charge. For example, an electron density of $n = 1.6 \times 10^{12}$ cm$^{-2}$ in the InSb nanosheet can be obtained at $V_{\text{BG}} = 4$ V. By normalizing the thickness of InSb nanosheet, a three-dimensional (3D) electron density of $n_{3D} = 8 \times 10^{17}$ cm$^{-3}$ can be obtained, yielding a Fermi wavelength $\lambda_F = 2(\frac{\pi}{3n_{3D}})^{\frac{1}{3}} \approx 22$ nm at $V_{\text{BG}} = 4$ V. This normalized bulk Fermi wavelength is larger than the nanosheet thickness (~20 nm) and thus the electron transport in the InSb nanosheet is of a 2D nature as it is assumed above. The electron mean free path in the nanosheet can be obtained from $L_e = \frac{\hbar \mu}{e}\sqrt{2\pi n}$, where $\hbar$ is the reduced Planck constant. At $V_{\text{BG}} = 4$ V, $L_e \sim 152$ nm is obtained, which is several times smaller than the channel length ($L \sim 600$ nm) and thus the carrier transport in the InSb nanosheet is in the diffusive regime. The inset of Fig. 2(d) shows a logarithmic scale plot of the $G - V_{\text{BG}}$ curves at $T = 1.9$ K. The black and red curves in the inset represent the results measured by sweeping $V_{\text{BG}}$ upward and downward, respectively, showing a clear ambipolar transport characteristic. The device exhibits a gate hysteresis of ~ 0.05 V at a sweeping rate of 0.08 V/s, which outperforms the previous devices made from free-standing InSb nanosheets directly on a SiO$_2$/Si substrate [22]. The smaller hysteresis seen here indicates that the influence of the charge traps in the SiO$_2$ dielectric has been effectively screened out by the graphite bottom-gate electrode and there are a significantly less amount of trapped charges in the hBN dielectric layer [32,44]. Again, it should be noted that a noticeably large gate hysteresis was observed at elevated temperatures (see Supplementary Information). For example, a gate hysteresis of ~0.56 V at 100 K and a gate hysteresis of ~1.16 V at 250 K were observed in the device at the same sweeping rate of 0.08 V/s. This increase in gate hysteresis at a high temperature is believed to originate dominantly from charge trapping and de-trapping in the bottom native oxidized layer of the InSb nanosheet [47-49].

Figure 2(e) shows the gate-transfer characteristics of the device measured at different temperatures. Here, no conductance quantization is observed in these measurements, which is simply due to the fact that the InSb nanosheet in the device is in the diffusive regime as it was shown above. Nevertheless, it is seen that at high temperatures (above ~100 K), the threshold voltage moves toward negative values with increasing temperature. This is consistent with the fact that the carrier density in the InSb nanosheet channel becomes higher due to thermal excitation at a higher temperature. However, the on-state conductance (taken at a positive $V_{\text{BG}}$ of above ~1 V) of the device is seen to decrease with increasing temperature at temperatures of above ~100 K. This is because of an increase in phonon scattering with increasing temperature. Figure 2(f) shows the field-effect mobilities extracted from the gate-transfer characteristic measurements of the device at different temperatures. It is seen that the mobility tends to saturate at low temperatures (below ~77 K), in good agreement with the results obtained from InSb QWs [15], but decreases exponentially with increasing temperature at high temperatures (above ~100 K). In the considered temperature range, the temperature-dependent mobility $\mu(T)$ in the InSb nanosheet can be approximately expressed as,

$$\frac{1}{\mu(T)} = \frac{1}{\mu_{\text{imp}}} + \frac{1}{\mu_{\text{ph}}(T)}. \tag{3}$$

Here, term $1/\mu_{\text{imp}}$ describes contributions from all scattering events due to imperfections, such as impurities, defects, charge traps, surface/interface roughness etc., in the nanosheet, its surface native oxides and the dielectrics [15,50]. This term is, to the lowest order approximation, temperature independent. Term $1/\mu_{\text{ph}}$ describes temperature-dependent contributions from acoustic and optical phonon scattering events. The mobility data extracted at different temperatures shown in Fig. 2(f) were fitted using Eq. (3) and the result of the fit is presented by the orange solid line in Fig. 2(f). The fit yields a value of $\mu_{\text{imp}}$~7900 cm² V⁻¹ s⁻¹ at the low temperature limit. At high temperatures (above ~100 K), the mobility is well fitted by a power law $\mu \sim T^{-\gamma}$ with $\gamma$~2.11 as indicated by the black dashed line in Fig. 2(f). This power-law fit confirms that optical phonon scattering was the dominant factor for the limitation of the electron mobility in the nanosheet at the high temperatures [51,52]. Note that acoustic phonon scattering would yield a $\mu \sim T^{-1}$ temperature dependence [53], which is not the case seen in Fig. 2(f).

*3.2. Electrical and magnetotransport measurements of the double-gate Hall-bar device*

Figure 3 shows transport measurements of the double-gate Hall-bar device at $T = 1.9$ K. Figures 3(a) and 3(b) show an optical image of an InSb nanosheet/hBN/graphite trilayer heterostructure (before fabricating small markers) and an SEM image of the double-gate Hall-bar device made from the trilayer heterostructure, while Fig. 3(c) shows a schematic view of the layer structure of the device and the circuit setup for the application of top and bottom gate voltages $V_{\text{TG}}$ and $V_{\text{BG}}$. In Fig. 3(a), two small, thick hBN pieces (marked by two black dotted lines), a large thin hBN piece (marked by a white dotted line), and a large graphite piece (marked by a purple dotted line) are seen. The thin hBN piece has a thickness of $t$~30 nm and has a large portion stacked on the graphite layer. The InSb nanosheet is seen to locate in the area marked by a black square on top of the stacked thin hBN/graphite bilayer. This InSb nanosheet has a width of $W$~ 900 nm and a thickness of ~30 nm (estimated based on the calibrated contrast in the SEM image) and is contacted by six Ti/Au electrodes as shown in Fig. 3(b). The source and drain electrodes 1 and 6 are ~1.7 μm apart, while the separations between probe electrodes 2 and 3 and between probe electrodes 3 and 5 are $L_{23}$~850 nm and $W_{35}$~500 nm, respectively. Note that the top gate and top gate dielectric cover almost the entire range of Fig. 3(b).

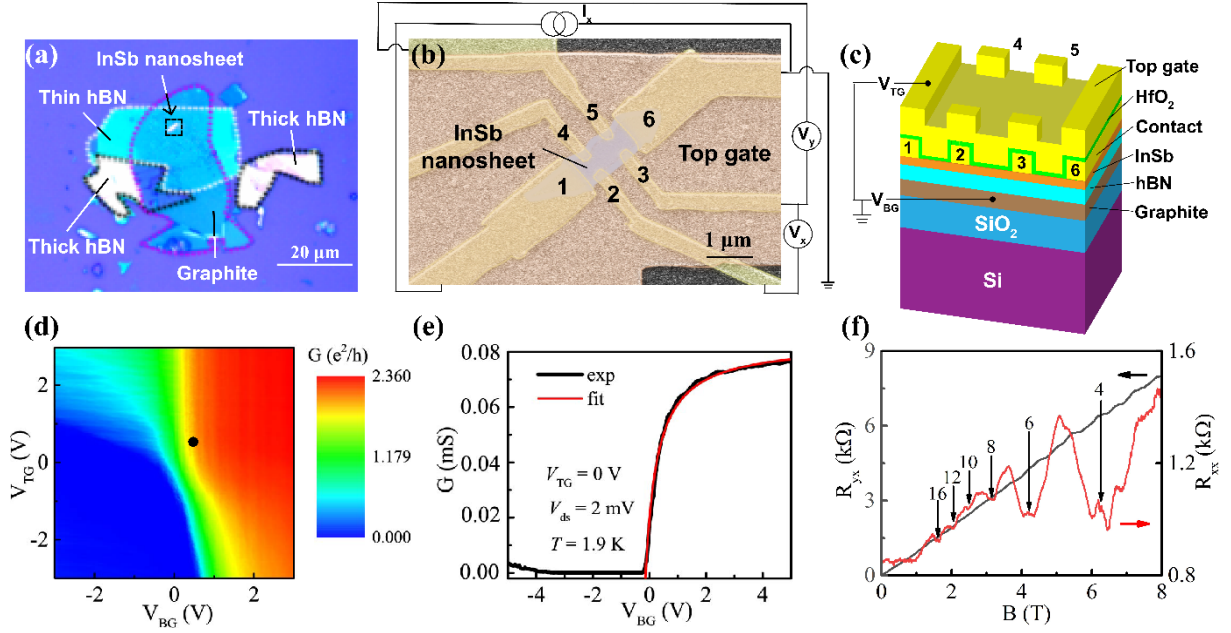

**Figure 3.** Electrical and magnetotransport measurements of a double-gate Hall-bar device at 1.9 K. (a) Optical image of the InSb nanosheet/hBN/graphite trilayer used for fabrication of the device. The InSb nanosheet is located in the area indicated by a black square. White, black, and purple dotted lines mark the thin hBN, thick hBN, and graphite regions in the sample, respectively. (b) False-colored SEM image of the device and magnetotransport measurement circuit setup. The nanosheet has a width of ~900 nm and a thickness of ~30 nm. The distance between electrodes 1 and 6 is ~1.7 μm, while separation between probe electrodes 2 and 3 is ~850 nm and that between probe electrodes 3 and 5 is ~500 nm. (c) Schematic view of the layer structure of the device and the circuit setup for the application of gate voltages $V_{TG}$ and $V_{BG}$. (d) Gate-transfer characteristics of the device measured in a two-terminal configuration by applying a DC voltage of $V_{ds}$=2 mV to electrodes 1 and 6 and recording the current through the InSb nanosheet channel $I_{ds}$ (with the conductance obtained through $G = I_{ds}/V_{ds}$) as a function of $V_{TG}$ and $V_{BG}$. (e) Measured $G$ of the device as a function of $V_{BG}$ at $V_{TG} = 0$ V (black curve). The red curve shows the fit of the on-state part of the measured data to Eqs. (1) and (2). (f) Measured $R_{xx}$ (red) and $R_{yx}$ (black) of the device as a function of magnetic field $B$ at $V_{BG}$=0.5 V and $V_{TG}$=0.5 V, i.e., a point marked by a black dot in (d). The measurements show well-defined Shubnikov-de Haas oscillations. The arrows mark the magnetic field positions for a few even values of filling factor $\nu$.

Figure 3(d) shows the gate-transfer characteristics of the double-gate device measured in a two-terminal configuration by applying a DC voltage ($V_{ds}$) of 2 mV to the source and drain electrodes (electrodes 1 and 6), i.e., the source-drain current $I_{ds}$ as a function of the voltages applied to the bottom and top gates $V_{BG}$ and $V_{TG}$ [see the circuit setup for $V_{BG}$ and $V_{TG}$ in Fig. 3(c)]. Owing to the short distances of the top and bottom gates to the InSb nanosheet, both gates show strong capacitive coupling to the nanosheet conduction channel. Figure 3(e) shows the measured two-terminal conductance $G = I_{ds}/V_{ds}$ of the InSb nanosheet as a function of $V_{BG}$ at $V_{TG} = 0$ V (black curve). The red curve in Fig. 3(e) shows the result of the fit of the measured on-state transfer characteristic data to Eqs. (1) and (2). With use of the channel length $L = 1.7$ μm, the channel width $W = 900$ nm, and the hBN thickness $t = 30$ nm, a field-effect mobility of $\mu_{FE} \sim 7200$ cm$^2$ V$^{-1}$ s$^{-1}$ can be extracted for the InSb nanosheet from the fit. Note that the extracted bottom-gate threshold voltage from the fit is slightly negative, which is different from the extracted value for the single bottom gate device as shown in Fig. 2(d). This difference is most likely due to the fact that the nanosheet in the double-gate Hall-bar device is about 50% thicker than the nanosheet in the single bottom-gate device and thus the conduction band in the nanosheet of the double gate device is less pushed up in energy by quantum confinement and the fact that the top gate voltage has been set to zero giving a heterostructure band alignment favorable for conduction channel opening at zero bottom-gate voltage.

Figure 3(f) shows the measured $R_{xx}$ (red) and $R_{yx}$ (black) as a function of magnetic field $B$ with both top and bottom gate voltages set at 0.5 V [i.e., at a point indicated by a black dot in Fig. 3(d)]. From the measured $R_{yx} - B$ curve at low magnetic fields, the Hall coefficient $R_H \sim 1030$ Ω/T and the sheet electron density $n_{sheet} \sim 6.1 \times 10^{11}$ cm$^{-2}$ in the InSb nanosheet can be obtained. The electron Hall mobility $\mu_{Hall}$ in the nanosheet can be estimated from the measured longitudinal resistance $R_{xx}$ at zero magnetic field and the sheet electron density $n_{sheet}$ via,

$$\mu_{Hall} = \frac{L_{23}}{W} \frac{1}{R_{xx} n_{sheet} e}. \qquad (4)$$

Using $L_{23} \sim 850$ nm, $W \sim 900$ nm, the zero-field value of $R_{xx} \sim 850$ Ω, and the sheet electron density $n_{sheet} \sim 6.1 \times 10^{11}$ cm$^{-2}$, a Hall mobility value of $\mu_{Hall} \sim 1.1 \times 10^4$ cm$^2$ V$^{-1}$ s$^{-1}$ is obtained. Note that if $W_{35} \sim 500$ nm is used instead of $W \sim 900$ nm in the calculation, a higher value of $\mu_{Hall} \sim 2.0 \times 10^4$ cm$^2$ V$^{-1}$ s$^{-1}$ would be obtained. The actual sheet Hall mobility at $V_{BG}=0.5$ V and $V_{TG}=0.5$ V would be a value in between the two above values. The measured Hall resistance $R_{yx}$ at high fields displays step-like structures, indicating appearance of the quantum Hall effect as a result of the formation of the Landau levels (LLs). In the same time, the longitudinal resistance $R_{xx}$ at high fields displays several pronounced peaks and valleys, i.e., well-defined Shubnikov-de Haas oscillations. The Shubnikov-de Haas oscillation peaks appear at the magnetic fields where the LLs coincide with the Fermi energy. Because each spin-split LL is highly degenerate with a unit area degeneracy of $eB/h$ (where $h$ is the Planck constant), the filling factor (FF), i.e., the number of spin-split LLs filled by electrons at each oscillation valley is given by $\nu = n_{sheet} h/eB$. The black arrows in Fig. 3(f) mark the magnetic field positions in the Shubnikov-de Haas oscillation valleys at which the filling factors $\nu$ at a few even values can be identified.

## 4. Conclusions

In conclusion, a method of fabrication of micrometer-sized InSb nanosheet/hBN/graphite trilayer heterostructures has been developed, and single- and double-gate devices have been made from the fabricated trilayer heterostructures on $SiO_2$/Si substrates. The method employs duplicated patterns on the transparent stacking/transfer stamps for position determination of hardly visible InSb nanosheets under the optical microscope and, thus, enables stacking of micrometer or smaller InSb nanosheets on exfoliated hBN flakes and on exfoliated graphite flakes. Based on the fabricated InSb nanosheet/hBN/graphite trilayer heterostructures, a single-gate InSb nanosheet field-effect device and a double-gate InSb nanosheet Hall-bar device are fabricated on $SiO_2$/Si substrates, where the hBN flakes are used as the bottom gate dielectrics and the graphite flakes as bottom gates. The fabricated devices are characterized by electrical and magnetotransport measurements. The measurements show that the InSb nanosheet in the single-gate field-effect device has an electron field-effect mobility of ~7300 $cm^2$ $V^{-1}$ $s^{-1}$ and a low gate hysteresis of ~0.05 V at 1.9 K, demonstrating an effective screening of charge traps in the $SiO_2$ dielectrics by the graphite gate. The double-gate Hall-bar device exhibits strong capacitive couplings of the two gates to the InSb nanosheet channel, and a Hall mobility of larger than $1.1 \times 10^4$ $cm^2$ $V^{-1}$ $s^{-1}$ as well as well-defined Shubnikov-de Haas oscillations at the sheet electron density of $6.1 \times 10^{11}$ $cm^{-2}$ and 1.9 K. The fabrication method established in this work should have the potential to be widely used in assembling heterostructures from many tiny sized planar materials for applications in nanoelectronics, spintronics, and quantum electronics.

## Acknowledgments

This work has been supported by the Ministry of Science and Technology of China through the National Key Research and Development Program of China (Grant Nos. 2017YFA0303304 and 2017YFA0204901), the National Natural Science Foundation of China (Grant Nos. 92165208, 92065106, 61974138, 11874071, 91221202, and 91421303), and the Beijing Academy of Quantum Information Sciences (Grant No. Y18G22). D. P. also acknowledges support from the Youth Innovation Promotion Association, Chinese Academy of Sciences (2017156).

**Data availability statement**

All data that support the findings of this study are included within the article and its Supplementary Information.

# Supplementary Information for
# Fabrication and characterization of InSb nanosheet/hBN/graphite heterostructure devices

Li Zhang[1], Yuanjie Chen[1], Dong Pan[2], Shaoyun Huang[1], Jianhua Zhao[2], and H. Q. Xu[1,3,*]

[1]*Beijing Key Laboratory of Quantum Devices, Key Laboratory for the Physics and Chemistry of Nanodevices, and School of Electronics, Peking University, Beijing 100871, China*

[2]*State Key Laboratory of Superlattices and Microstructures, Institute of Semiconductors, Chinese Academy of Sciences, P.O. Box 912, Beijing 100083, China*

[3]*Beijing Academy of Quantum Information Sciences, Beijing 100193, China*

[*]Corresponding author: hqxu@pku.edu.cn

(Date: April 18, 2022)

## Outline

S1. InSb nanosheet, hBN and graphite flake materials, and stamps

S2. Process details for fabrication of InSb nanosheet/hBN/graphite trilayer heterostructures

S3. Hysteresis in gate voltage in the single-gate field-effect device at different temperatures

**S1. InSb nanosheet, hBN and graphite flake materials, and stamps**

The planar InSb nanosheet/hexagonal boron nitride (hBN)/graphite trilayer devices studied in the main article were made from free-standing InSb nanosheet, thin hBN flakes and graphite flakes. The free-standing InSb nanosheets were grown on InAs nanowire stems on a Si(111) substrate by molecular beam epitaxy (MBE). The growth process was started by depositing a thin layer of Ag on the Si(111) substrate in a MBE chamber. The sample was then annealed in situ to generate Ag nanoparticles. Subsequently, InAs nanowires were grown on these Ag nanoparticles at a temperature of 505 °C for 20 min with the As/In beam equivalent pressure ratio being set at 30. After that, the group-V source was switched abruptly from As to Sb and free-standing InSb nanosheets were formed on top of InAs nanowires by gradually increasing the Sb flux while maintaining the In flux unchanged. As-grown InSb nanosheets were examined by scanning electron microscopy and transmission electron microscopy measurements. The measurements showed that the nanosheets were zincblende crystals, free from stacking faults and twin defects and were ~1 μm in lateral size and ~10-80 nm in thickness. The MBE grown InSb nanosheets were mechanically picked up using a dry clean-room paper and were then randomly distributed on the surface of a SiO$_2$/Si substrate with prefabricated, optically visible Ti/Au markers on top (sample substrate A). The hBN and graphite flakes were obtained by mechanical exfoliation from commercially available bulk hBN (from HQ Graphene) and graphite (from Graphene Supermarket) materials using Scotch tapes and were transferred on other two marker-prefabricated SiO$_2$/Si substrates (sample substrates B and C). Here, we note that before attaching the hBN- or graphite-loaded tapes to substrates B and C, the surfaces of the substrates were cleaned by oxygen plasma. While after the tapes were attached, the samples B and C were heated at 100 °C on a hot plate for a few minutes before tape separation in order to obtain flakes with large lateral sizes [1]. After tape separation, the entire samples B and C were dipped in acetone to remove the residues of the adhesive substances and organic contamination on the material surfaces.

In order to assemble prepared micrometer-sized InSb nanosheet on hBN flakes and then on graphite flakes to form InSb nanosheet/hBN/graphite trilayer heterostructures, transparent stamps were also needed. In the work presented in the main article, the transparent stamps were made first by spin coating of polypropylene carbonate (PPC) (from Sigma Aldrich, 20% concentration dissolved in tetrahydrofuran) onto a polydimethylsiloxane (PDMS) sheet [2]. The PPC/PDMS layer was then baked at 90 °C for 10 min and was cut into small stamps of about 3 mm × 3 mm in size.

**S2. Process details for fabrication of InSb nanosheet/hBN/graphite trilayer heterostructures**

The fabrication of InSb nanosheet/hBN/graphite trilayer heterostructures was carried out using a home-made alignment stacking/transfer system equipped with an optical microscope, a micro-positioner, a sample stage (from Instec) with a vacuum chuck and a temperature controller. Figure S1 shows the schematics for the process of fabrication of an InSb nanosheet/hBN/graphite trilayer heterostructure. First, as shown in the bottom graph of Fig. S1(a), a PPC/PDMS stamp was placed onto a glass microscope slide. The glass microscope slide was then flipped upside down and was mounted on the micro-positioner with the PPC film facing down, as shown in the top graph of Fig. S1(a). Next, sample substrate A was placed on the sample stage and was fixed firmly via the vacuum chuck, and the region containing markers was brought into focus in the field of view of the optical microscope as shown in Fig. S1(b). Here and after, for clarity, we only illustrate a magnified view of the region, where the target materials were seen and to be positioned, aligned, captured, and/or released, in Figs. S1(b)-S1(i). The middle graph of Fig. S1(b) shows schematically a region of the glass microscope slide with a part of the PDMS/PPC stamp that was used to pick up materials. By adjusting the micro-positioner, the stamp was coarsely positioned over an area of sample substrate A, where an InSb nanosheet located beside marker '11' was found, see Fig. S1(b). After aligning the stamp with the area on sample substrate A, the stamp was lowered until the PPC film touched the surface of sample substrate A, as shown in Fig. S1(c). Then, the substrate A was heated up through heating the sample stage to a temperature of 35-40 °C, a temperature at which the PPC film had transformed from a rigid glassy state to a soft rubberlike state [3,4] and became ductile and easily deformable. The tight contact between the PPC film and sample substrate A was maintained by pressing the stamp against sample substrate A. After 3-5 min, the stamp with adhesion of the InSb nanosheet on the PPC surface was lifted from sample substrate A [see Fig. S1(d)], and the sample stage, sample substrate A and the PDMS/PPC stamp were cooled down to room temperature. After being cooled down in temperature, the PPC film returned to the glassy state with clear patterns of the markers on substrate A being duplicated on the PPC surface [Fig. S1(d)].

To assemble the picked-up InSb nanosheet on an exfoliated hBN flake, sample substrate A was replaced with sample substrate B on the sample stage and the region containing the hBN flake was brought into focus in the field of view of the optical microscope. Here, the key step of assembling the InSb nanosheet on the hBN flake is an optical alignment of the InSb nanosheet to the hBN flake. However, since the size of the picked-up InSb nanosheet was at a micrometer or sub-micrometer scale and the image seen through the stamp was not sharp, the captured InSb nanosheet appeared to be a faint speck submerging in a vast blurred background, making the alignment process difficult. To make the alignment efficiently, the duplicated patterns on the PPC surface, which had a size of about 8 micrometers, were located in the following fabrication step. As the stamp was lowered closer and closer to sample substrate B, the duplicated patterns on the PPC surface became more and more focused [see Fig. S1(e)] and the region beside pattern '11' was aligned with the hBN flake by fine adjusting the positions of the micro-positioner and the sample stage. Once aligned, the PPC film on the stamp was brought into contact with the surface of sample substrate B, as shown in Fig. S1(f). Then, substrate B was heated to a temperature of 35-40 °C and the PPC film maintained tight contact to the surface of sample substrate B by keeping the stamp pressed against sample substrate B. The PPC film was converted to the rubberlike state again and deformed within a couple of seconds after the temperature rise, as demonstrated by evanescence of the duplicated patterns on the PPC film [Fig. S1(f)]. After 5 min, the stamp with adhesion of the InSb nanosheet/hBN bilayer on the PPC surface was lifted from sample substrate B [see the upper graph in Fig. S1(g)], and the sample stage, sample substrate B and the stamp were cooled down to room temperature again.

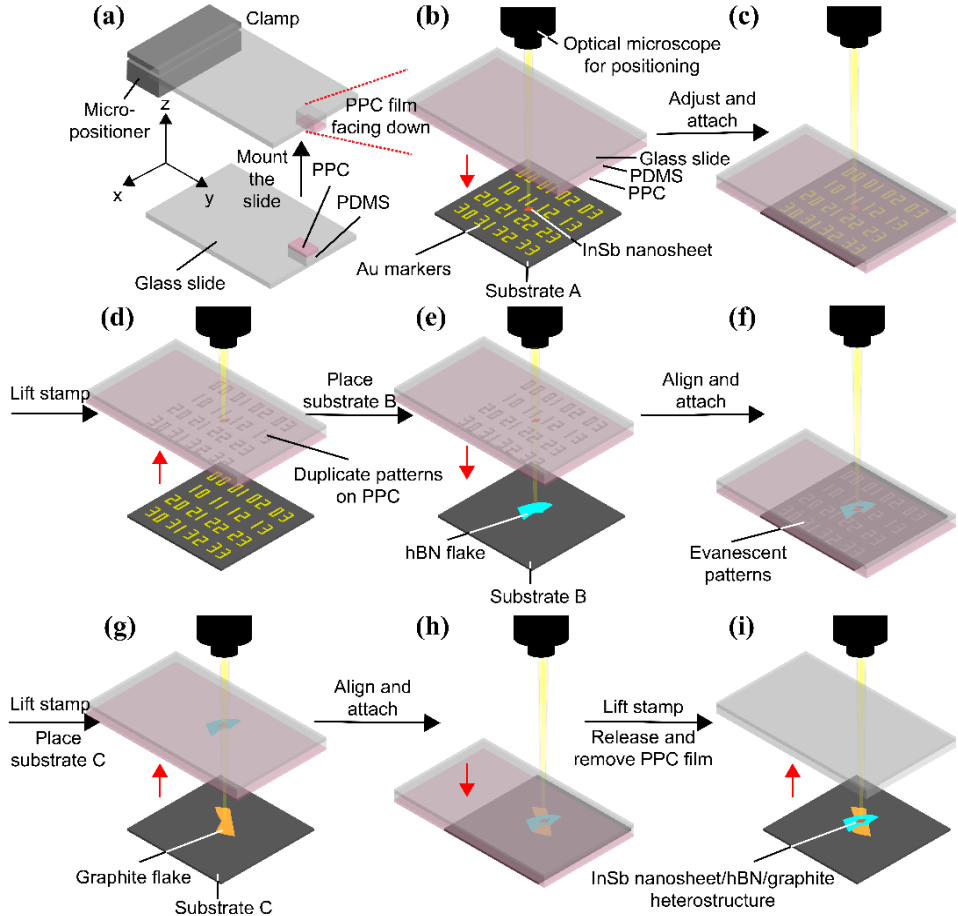

**FIG. S1.** Schematic process flow of an InSb nanosheet/hBN/graphite trilayer heterostructure fabrication. (a) Bottom graph: a PPC/PDMS stamp was placed on the glass microscope slide. Top graph: the glass microscope slide was clamped on the micro-positioner. (b) Top graph: optical microscope. Middle graph: magnified view of the region of the glass microscope slide with the PDMS/PPC stamp. Bottom graph: sample substrate A, with a micrometer-sized InSb nanosheet and predefined markers, on the sample stage (not shown). (c) The PPC film in tight contact to sample substrate A. (d) The stamp with picked up InSb nanosheet was lifted from sample substrate A. Here a visible pattern of the substrate markers duplicated on the PPC film is also indicated. (e) Sample substrate A was replaced with sample substrate B. The stamp was lowered and the region beside the duplicated pattern '11' was aligned with a target hBN flake on sample substrate B. (f) The PPC film was in contact to sample substrate B. (g) The target hBN flake had been picked up, leading to the formation of the InSb nanosheet/hBN bilayer on the PPC surface. Then sample substrate B was replaced by sample substrate C and a graphite flake on substrate C was aligned with the InSb nanosheet/hBN bilayer. (h) The PPC film was in contact to sample substrate C. (i) The PPC film was released from the PDMS and the PPC film/InSb nanosheet/hBN structure was stacked onto the graphite flake. Later the sample was dipped in acetone to remove the PPC film, leaving a micrometer-sized InSb nanosheet/hBN/graphite trilayer heterostructure on substrate C.

Afterwards, sample substrate B was replaced with sample substrate C on the sample stage [see lower graph in Fig. S1(g)] and the region containing the target graphite flake on sample substrate C was brought into focus in the field of view of the optical microscope. Then, the stamp was lowered and the InSb nanosheet/hBN bilayer was aligned with the graphite flake, with the help of the large sized hBN flake, by adjusting the positions of the micro-positioner and the sample stage. After the PPC film on the stamp was brought into contact with the surface of sample substrate C [Fig. S1(h)], the substrate C was heated to a temperature of 120 °C. After 10 min, the stamp was lifted from sample substrate C with the PPC film being released from the PDMS and the PPC film/InSb nanosheet/hBN structure being stacked onto the graphite flake. Finally, the sample was dipped in acetone to remove the PPC film, leaving an InSb nanosheet/hBN/graphite trilayer heterostructure on sample substrate C, as shown in Fig. S1(i).

**S3. Hysteresis in gate voltage in the single-gate field-effect device at different temperatures**

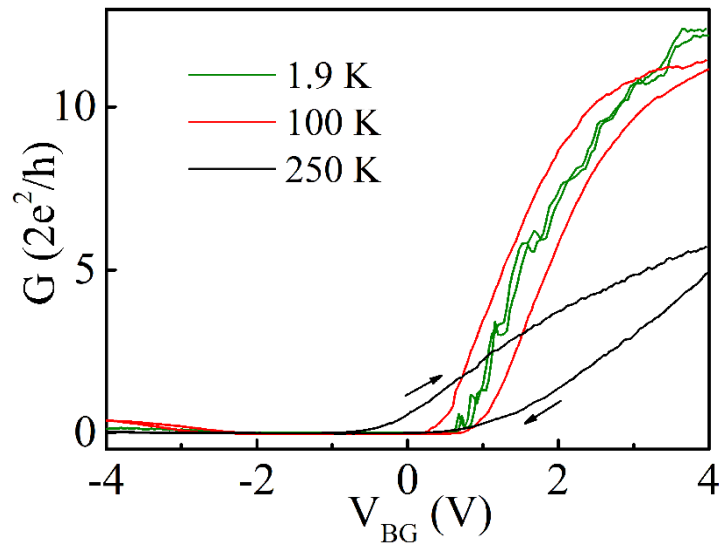

**FIG. S2.** The transfer characteristic curves of the InSb nanosheet/hBN/graphite heterostructure single-gate field-effect device measured at different temperatures. The hysteresis increases with increasing temperature. The directions of gate voltage sweeping are indicated by the arrows.

In this section, the observation of the gate hysteresis in the gate-transfer characteristic measurements of the single-gate field-effect device as studied in the main article at high temperatures is briefly presented. Figure S2 shows the gate-transfer characteristic curves of the single-gate field-effect device measured at temperatures of $T$ = 1.9, 100 and 250 K with the gate voltage being swept in both upward and downward directions. Here the measurement data at 1.9 K are the same as those shown in Fig. 2(d) of the main article. It is seen that the gate hysteresis in the measured gate-transfer characteristic curve at 1.9 K is ~0.05 V, while gate hysteresis becomes ~0.56 V at 100 K, and ~1.16 V at 250 K (i.e., gate hysteresis becomes larger at a higher temperature. Similar gate hysteresis has been observed in the devices made with an oxidized layer as a gate dielectric and has been attributed dominantly to charge trapping and de-trapping processes in the oxide dielectric layer [5-7]. Here, the gate hysteresis observed in the InSb nanosheet/hBN/graphite trilayer single-gate field-effect device is primarily due to the charge trapping and de-trapping processes in the InSb native oxidized layer located between the InSb nanosheet and the hBN flake. Such charge trapping and de-trapping processes become more and more active with increasing temperature, leading a larger gate hysteresis at a higher temperature in consistence with the observation shown in Fig. S2.